\documentclass[aps,prb,twocolumn,showpacs]{revtex4}
\usepackage{tabularx,graphicx}

\begin{document}

\title{Quantum Phase Transitions in Dissipative Tunnel Junctions}
\author{Scott Drewes\cite{sdloc}, Daniel P. Arovas\cite{dpa}, and Scot Renn\cite{srloc}}
\affiliation{
Department of Physics, University of California at San Diego,
La Jolla CA 92093\\
}

\date{\today}

\begin{abstract}
The Ueda-Guinea model of a dissipative tunnel junction is investigated.
This model accounts for final state effects associated with single-electron
tunneling.  A quantum phase transition emerges, marking a boundary between
insulating (Coulomb blockade) and conducting phases.  The system is analyzed
by large-$N$ techniques, self-consistent harmonic approximation, and
Monte Carlo methods.\end{abstract}
\maketitle

\section{Introduction}

Over 30 years ago, Mahan\cite{m67} and Nozieres and deDominicis\cite{nd69} predicted the existence of power law behavior in the absorption edges for X-ray transitions in metals.  This
phenomenon is due to the influence of a suddenly switched-on potential, due to the (screened) core hole, on the electrons.  There are two aspects to this physics.  One is the orthogonality 
catastrophe\cite{pwa67} due to the change of electronic wavefunctions in the presence of
the core hole.  The other is formation of an excitonic resonance between the liberated electron
and the core hole\cite{m67,nd69}.  Recently, several authors have noted the relevance of such 
nonequilibrium effects to mesoscopic systems such as tunnel junctions\cite{ug91,ml92} and quantum dots\cite{mgb96,siv96,agam97}.

The model we will study has been described in the recent work of Bascones {\it et al.}\cite{bas00},
based on the original work of Ueda and Guinea\cite{ug91}.  Briefly, we consider a junction 
consisting of two banks, left (L) and right (R), each described by a noninteracting Hamiltonian
of the form $ {\cal H}_\alpha=\sum_ {\bf k}  \varepsilon ^{\vphantom{\dagger}}_\alpha( {\bf k})\,c ^\dagger_{ {\bf k}\alpha} c ^{\vphantom{\dagger}}_{ {\bf k}\alpha}$,
where $\alpha= {\rm L}, {\rm R}$.  The tunneling between these banks is described by
$ {\cal H}_{\rm tunnel}=-t\,e^{i\phi}\sum_{ {\bf k}, {\bf k}'}c ^\dagger_{ {\bf k} {\rm R}}c ^{\vphantom{\dagger}}_{ {\bf k}' {\rm L}}+{\rm H.c.}$
The phase variable $\phi$ is a collective coordinate conjugate to the charge transfer $Q$
between left and right banks: $[Q,e^{i\phi}]=e^{i\phi}$.  Associated with this charge transfer
is a Coulomb energy, $ {\cal H}_Q=(Q-Q_{\rm offset})^2/2C$, where $C$ is a capacitance and
$Q_{\rm offset}\equiv \nu e$ accounts for the charge environment of the junction.  Finally,
the interaction between electrons and the global charge is written as\cite{ug91,bas00,drg98}
\begin{equation}
 {\cal H}_{\rm int}=(Q-Q_{\rm offset})\sum_{ {\bf k}, {\bf k}',\alpha} U^{\alpha\vphantom{\dagger}}_{ {\bf k} {\bf k}'}\,
c ^\dagger_{ {\bf k}\alpha}\,c ^{\vphantom{\dagger}}_{ {\bf k}'\alpha}\ .
\end{equation}

The fermionic degrees of freedom are quadratic in the Hamiltonian $ {\cal H}= {\cal H}_ {\rm L}+ {\cal H}_ {\rm R}
+ {\cal H}_{\rm tunnel}+ {\cal H}_Q+ {\cal H}_{\rm int}$, and can be integrated out\cite{bms83}, leaving an effective model whose only dynamical degree of freedom is the phase field $\phi$.  When $U^\alpha_{ {\bf k} {\bf k}'}=0$, the resultant effective action is that obtained by Ben-Jacob, Mottola,
and Sch{\"o}n (BMS)\cite{bms83}.  This approach implicitly assumes that the energy level spacing
in each of the banks is small on the scale of the charging energy $E_ {\rm c}\equiv e^2/2C$
and temperature $ k_{\scriptscriptstyle {\rm B}}T$.  When the effects of $ {\cal H}_{\rm int}$ are accounted for, one obtains
a modified Euclidean BMS action of the form\cite{ug91,bas00}
\begin{eqnarray}
S[\phi(s)]&=&{1\over 4} \int\limits_0^L\!\! ds
\left({ {\partial}\phi\over {\partial} s}\right)^2 +\alpha \int\limits_0^L\!\! ds  \int\limits_0^L\!\! ds'\, K(s-s')\nonumber\\
&&\qquad\times\Big\{1-\cos\Big(\phi (s)-\phi (s')\Big)\Big\}\ .
\label{eact}
\end{eqnarray}
The kernel $K(s-s')$ is given by
\begin{equation}
K(s)=\left[{\pi\over L}\,\csc\left({\pi |s|\over L}\right)\right]^{2- \epsilon}\ .
\end{equation}
Here, $L=E_ {\rm c}/ k_{\scriptscriptstyle {\rm B}}T$ is the dimensionless inverse temperature, and $\alpha=g_\infty/4\pi^2$,
where $g ^{\vphantom{\dagger}}_\infty$ is the high temperature conductance of the junction\cite{htc} in units
of $e^2/h$.  The BMS model is recovered for $ \epsilon=0$.  The parameter $ \epsilon$ is a sum
over contributions from the two banks $j= {\rm L}, {\rm R}$, with $ \epsilon_j=-(\delta_j/\pi)^2$
due to the orthogonality catastrophe and $ \epsilon_j=(2\delta/\pi)-(\delta_j/\pi)^2$ if excitonic
effects are relevant.

When the tunnel junction is placed in series with a capacitor, forming a single electron box
\cite{sct92}, the external charge environment is accounted for by a topological term in the
action (\ref{eact}), $\Delta S_{\rm top}=-2\pi i \nu\,W[\phi]$, where $W[\phi]=[\phi(L)-\phi(0)]/2\pi$
is the winding number of the phase field $\phi$.

The model is a version of the ubiquitous dissipative quantum rotor\cite{sz90}.
The $ \epsilon=0$ case has been studied by several authors\cite{zs91,hsz99,fsz95,gop98,weg97}.
Physical quantities are smoothly dependent on $\alpha$ and there is no phase
transition at any finite $\alpha$\cite{sca91}.  For $ \epsilon>0$, however, a quantum ($T=0$) phase
transition is present at a critical value $\alpha_ {\rm c} ( \epsilon)$, as first noted by Kosterlitz\cite{kos77}.
Here, we investigate this phase transition using a large-$N$ expansion, the self-consistent
harmonic approximation, and finally Monte Carlo simulations.

\section{Large-$N$ Theory}\label{SlargeN}

The large-$N$ generalization of this problem was first discussed in unpublished work
by Renn\cite{renn97}.  Consider the action
\begin{eqnarray}
&&S[ {\bf n} (s),\lambda(s)]= \int\limits_0^L\!\! ds\left\{{1\over 4}
\left({ {\partial} {\bf n}\over {\partial} s}\right)^2 + \lambda(s)\left( {\bf n}^2 (s)-qN\right)
\right\}\nonumber\\
&&\qquad+ \frac{1}{2}\,\alpha \int\limits_0^L\!\! ds  \int\limits_0^L\!\! ds'\, K(s-s')\,| {\bf n} (s)- {\bf n} (s')|^2\ ,
\end{eqnarray}
where $ {\bf n} (s)$ is a real $N$-component  vector.  The field $\lambda(s)$ serves as
a Lagrange multiplier which enforces the constraint $ {\bf n}^2 (s)=qN$ for all $s$; typically
we take $q=1/N$, so the $ {\bf n}$ field is of unit length.  When $N=2$, one can eliminate
the constraint with the parameterization $ {\bf n}=(\cos\phi,\sin\phi)$, whence one recovers
the action of eqn. (\ref{eact}).

In the $N\to\infty$ limit, the action becomes dominated by the saddle point $\lambda(s)=\lambda$,
a constant function.  One then has
\begin{eqnarray}
S&=&\sum_{\omega_n}\left[\frac{1}{4}\omega_n^2
+\lambda + \alpha\,[ {\hat K} (0)- {\hat K} (\omega_n)]\right]\,| {\hat{\bf n}} (\omega_n)|^2
\nonumber\\
&&\qquad-qNL\lambda\ ,
\end{eqnarray}
where $\omega_n=2\pi n/L$ is a bosonic Matsubara frequency, and
\begin{eqnarray}
 {\bf n} (s)&\equiv&{1\over\sqrt{L}}\sum_{\omega_n} {\hat{\bf n}} (\omega_n)
\,e^{-i\omega_n s}\\
 {\hat K} (\omega_n)&=& \int\limits_0^L\!\! ds\,K(s)\,e^{i\omega_n s}\ .
\end{eqnarray}
The saddle point theory is thus a Gaussian theory, with correlation functions
\begin{eqnarray}
\langle n_i(s)\,n_j(s')\rangle&=&G(s)\,\delta_{ij}\\
G(s)&=&{1\over 2L}\sum_{\omega_n} {e^{-i\omega_n s}\over
\frac{1}{4}\omega_n^2+\lambda + \alpha[ {\hat K} (0)- {\hat K} (\omega_n)]}\ .
\nonumber
\end{eqnarray}
Extremizing the free energy $F=-L^{-1} \mathop{\rm Tr}_ {\hat{\bf n}}\exp(-S[ {\hat{\bf n}}])$
with respect to $\lambda$ yields the equation $G(0)=q$, which is
to be solved for $\lambda(\alpha,L)$.

In the zero temperature ($L\to\infty$) limit, $\omega_n\to\omega$ becomes
a continuous quantity, and $ {\hat K} (0)- {\hat K} (\omega)=C_ \epsilon\,|\omega|^{1- \epsilon}$,
with
\begin{equation}
C_ \epsilon={\pi\over\Gamma(2- \epsilon)\,\cos( \frac{1}{2}\pi \epsilon)}\ .
\end{equation}
Straightforward analysis of the integral shows that $\lambda$ vanishes
for $\alpha> \alpha_ {\rm c} ( \epsilon)$, where
\begin{eqnarray}
 \alpha_ {\rm c} ( \epsilon)&=&{1\over 4C_ \epsilon}\left({2A_ \epsilon\over\pi q \epsilon}\right)^{1+ \epsilon}\\
A_ \epsilon&\equiv& \int\limits_0^\infty\!\!{dt\over 1+t^{1+ \epsilon^{-1}}}\ .
\end{eqnarray}
From $\lim_{ \epsilon\to 0} A_ \epsilon = 1$ and $C_{ \epsilon=0}=\pi$, we find,
restoring $q=1/N$, that for small $ \epsilon$ the critical point occurs at
$ \alpha_ {\rm c} ( \epsilon)=N/2\pi^2 \epsilon$.  This is to be contrasted with the renormalization
group result of Kosterlitz\cite{kos77}, $ \alpha_ {\rm c}^{\scriptscriptstyle{\rm RG}} ( \epsilon)=(N-1)/2\pi^2 \epsilon$.
Note that these results agree to leading order in $1/N$.

Further analysis reveals the critical behavior of $\lambda(\alpha)$ in
the vicinity of the critical point:
\begin{eqnarray}
\lambda(\alpha)&\simeq&D_ \epsilon\,\big( \alpha_ {\rm c} ( \epsilon)-\alpha\big)^\nu\\
D_ \epsilon&=&\big[ 2^{1- \epsilon}C_ \epsilon  \alpha_ {\rm c} ( \epsilon)\big]^{2\over 1+ \epsilon}\,
\left({A_ \epsilon/B_ \epsilon\over (1+ \epsilon)\, \alpha_ {\rm c} ( \epsilon)} \right ) ^\nu\\
B_ \epsilon&=& \int_{0}^{\infty}\!{dt\over 1+ t^{ \epsilon^{-1}-1}}\quad {\rm if}
\ 0\le \epsilon< \frac{1}{2}\nonumber\\
&=&{ \epsilon\over 2 \epsilon-1} \int_{0}^{\infty}\!{dt\over
\left(1+ t^{1+ \epsilon\over 2 \epsilon-1}\right)^2}\quad {\rm if}
\  \frac{1}{2}\le \epsilon\nonumber
\end{eqnarray}
where $\nu={\rm max}(1, \epsilon^{-1}-1)$.

\begin{figure} [t]
\includegraphics[width=9cm]{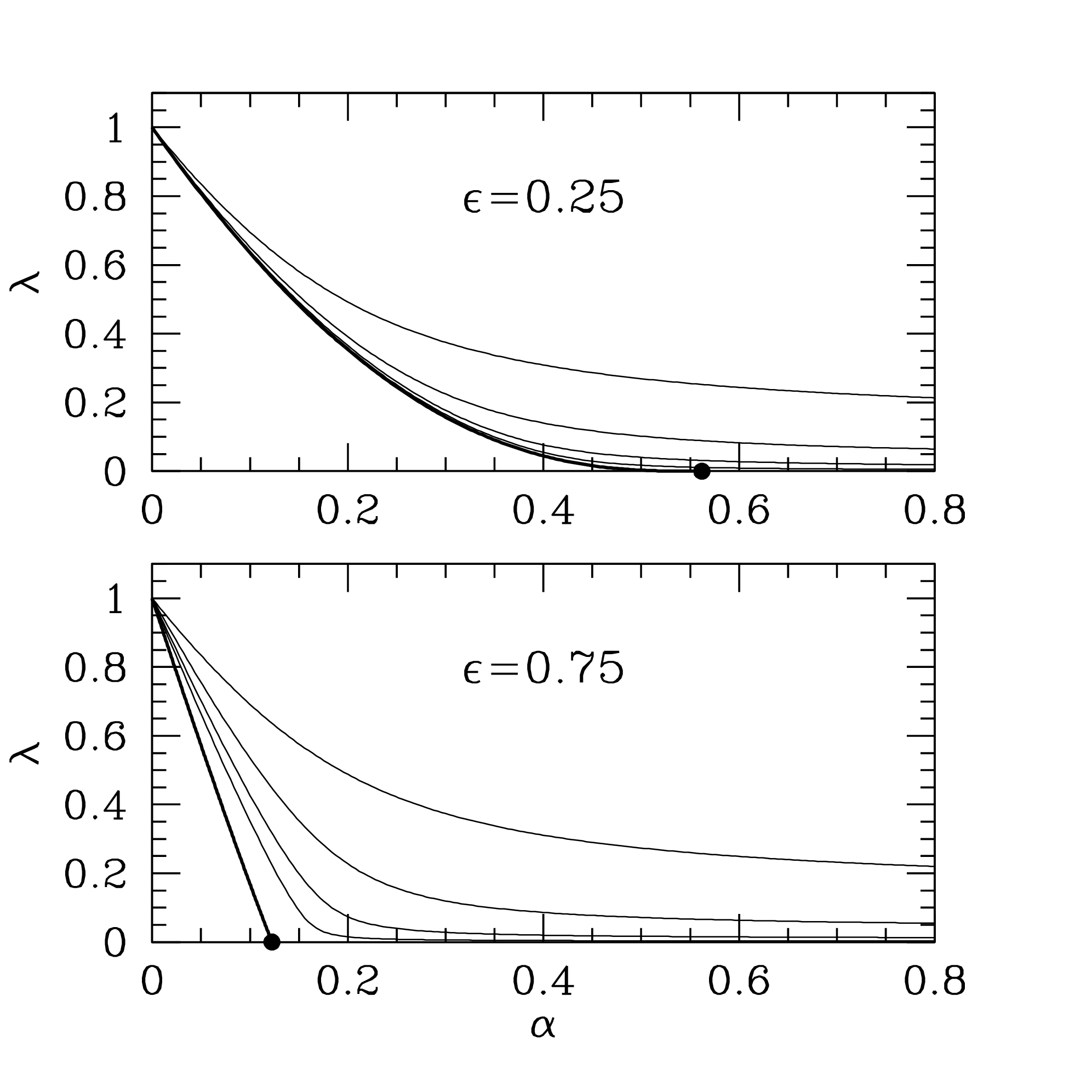}
\caption
{\label{lnc} Solution to the large-$N$ model at finite $L$ with $ \epsilon=0.25$ (top)
and $ \epsilon=0.75$ (bottom).  $\lambda$ {\it versus\/} $\alpha$ for $L=8$, $32$,
$128$, $512$ (thin lines) and $L=\infty$ (thick line).  $ \alpha_ {\rm c} ( \epsilon=0.25)=0.5619$,
and $ \alpha_ {\rm c} ( \epsilon=0.75)=0.1226$.}
\end{figure}

\subsection{Kubo Formula}

When $N=2$ the symmetry group $ {\rm O} (2)$ has a single generator, hence there
is one vector potential.  For the $ {\rm O} (N)$-symmetric case, a local gauge is
effected by $ {\bf n} (s)\to  {\cal R} (s)\, {\bf n} (s)$, where $ {\cal R} (s)\in  {\rm O} (N)$ is given by
\begin{equation}
 {\cal R} (s)=\exp\left[iA_a (s) {\rm T}^a\right]
\end{equation}
where the $ {\rm T}^a$ generate the Lie algebra $ {\rm o} (N)$.  These $ \frac{1}{2} N(N-1)$
generators are $N\times N$ Hermitian antisymmetric tensors; they may
be chosen to satisfy the normalization $ \mathop{\rm Tr} ( {\rm T}^a  {\rm T}^b)=N\,\delta^{ab}$.
(For $N=2$ the sole generator is $ {\rm T}=\sigma^y$.)
There are therefore $ \frac{1}{2} N(N-1)$ vector potentials $A_a(s)$, variation with
respect to which defines the $ {\rm O} (N)$ currents,
\begin{eqnarray}
I_ {\rm a} (s)&=&{\delta S_{\rm int}[ {\cal Y} (s) {\bf n} (s)]
\over\delta A_a (s)}\Big|_{ {\bf A}=0}\nonumber\\
&=&2i\alpha\, n_k (s)\,  {\rm T}^a_{kl} \int\limits_0^L\!\! ds\,K(u-s)\,n_l(s)
\end{eqnarray}
and the noise current-current correlation function
\begin{eqnarray}
&&\langle  {\cal T}\,I^ {\rm n}_a(s)\,I^ {\rm n}_b(0)\rangle =\Big\langle {\delta^2\! S_{\rm int}
\over\delta A_a(s)\,\delta A_b(0)}\Big\rangle_{ {\bf A}=0}\\
&&=2N\alpha \delta^{ab}\Big[\delta(s) \int\limits_0^L\!\! du\,K(u) G(u)-K(s)G(s)\Big]\ .
\nonumber
\end{eqnarray}
The conductance (in units of $e^2/h$) is computed according to the Kubo formula \cite{cfoot},
\begin{equation}
g_{ab}(i\omega_n)={2\pi\over\omega_n}\,
 \int\limits_0^L\!\! ds\,e^{i\omega_n s}\langle  {\cal T}\,I^ {\rm n}_a(s)\,I^ {\rm n}_b(0)\rangle\ .
\label{kubo}
\end{equation}
We define $g(i\omega_n)$ by $g_{ab}(i\omega_n)\equiv N g(i\omega_n)\,\delta_{ab}$.
We have no expression for the analytic continuation of the conductance to real
frequencies.   As a diagnostic of any phase transition, we will examine the quantity
$g_{\scriptscriptstyle{\rm A}}\equiv g(i\omega_{n=1})$.  We find
\begin{equation}
g_{\scriptscriptstyle{\rm A}}={4\pi^2\alpha\over L}\left({L\over\pi}\right)^ \epsilon\! \int\limits_0^L\!\! ds\,\sin^ \epsilon\left({\pi s/L}\right)
\big\langle\!\cos\big(\phi(s)-\phi(0)\big)\big\rangle\ .
\label{gaeqn}
\end{equation}
A similar expression was proposed by Bascones {\it et al.\/} \cite{bas00}:
\begin{equation}
g_{\scriptscriptstyle{\rm B}}=4\pi^2\alpha\,\left({L\over\pi}\right)^ \epsilon\,
\big\langle\!\cos\big(\phi( \frac{1}{2} L)-\phi(0)\big)\big\rangle\ .
\end{equation}

At the quantum critical point, we can evaluate DC
($\omega\to 0$) limit of $g(i\omega_n,\alpha_ {\rm c},T=0)$.  We obtain
\begin{equation}
g_ {\rm c}=\pi\,(1- \epsilon)\, \,{\rm ctn\,} ( \frac{1}{2}\pi \epsilon)\ .
\end{equation}
Note that while the critical coupling $\alpha_ {\rm c}$ is $q$-dependent, the
critical conductance $g_ {\rm c} ( \epsilon)$ is universal and independent of $q$.
The approximations $g_{\scriptscriptstyle\rm {A}}$ and $g_{\scriptscriptstyle{\rm B}}$ also take on universal values
at criticality:
\begin{eqnarray}
g_{\scriptscriptstyle{\rm A}}^ {\rm c} & = & {1\over\pi}\int\limits_0^\pi\! d\theta\,\left( {\sin\theta\over\theta} \right)^ \epsilon
\times g_ {\rm c} ( \epsilon) \\
g_{\scriptscriptstyle{\rm B}}^ {\rm c} & = & \left( {2\over\pi} \right)^{1+ \epsilon} \times g_ {\rm c} ( \epsilon)\ .
\end{eqnarray}

\begin{figure} 
\includegraphics[width=9cm]{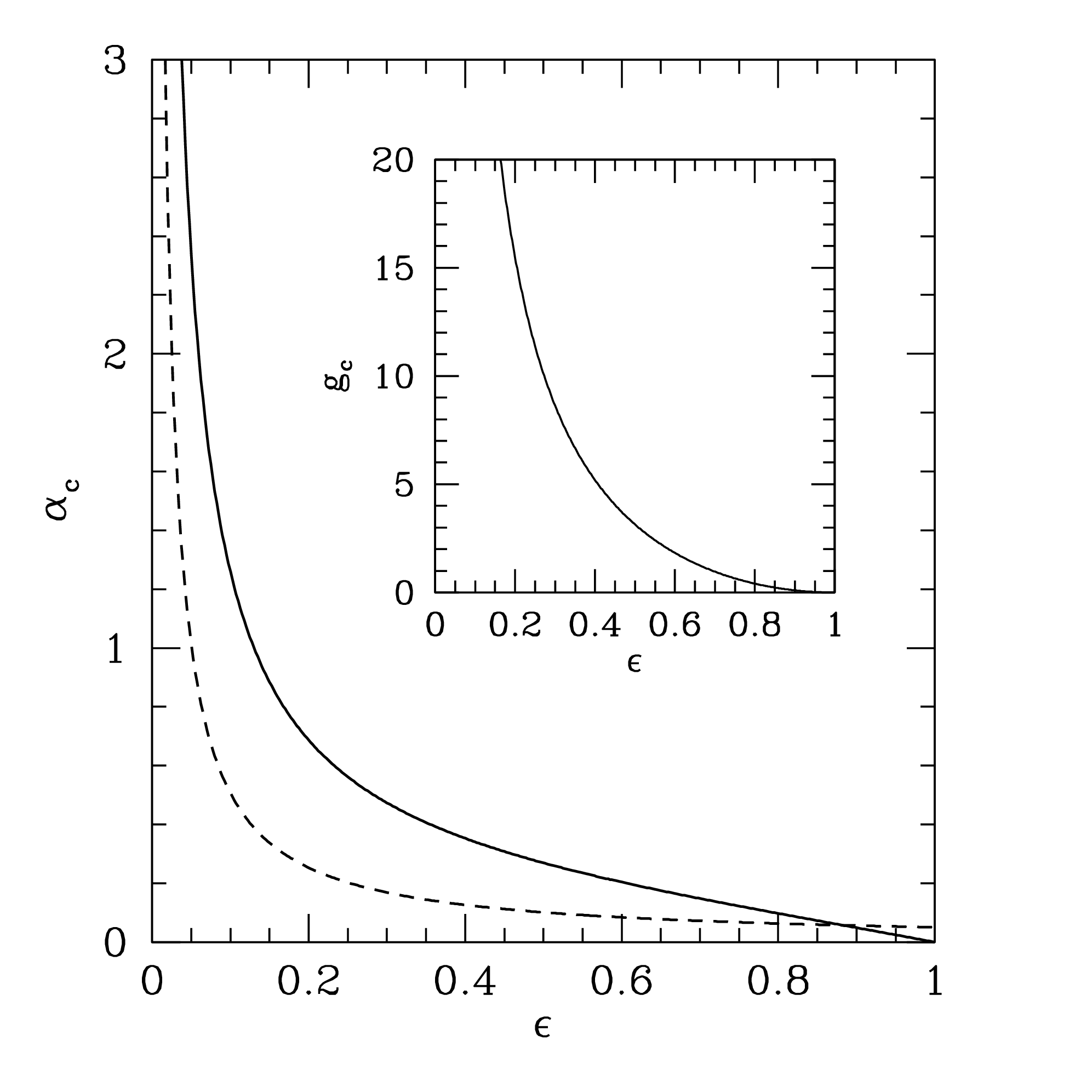}
\caption
{\label{nagc} Large-$N$ results for $ \alpha_ {\rm c} ( \epsilon)$ (solid) compared with 
the Kosterlitz RG value $ \alpha_ {\rm c}^{\scriptscriptstyle{\rm RG}}=1/2\pi^2 \epsilon$ (dashed).  Inset:
critical conductance $g_ {\rm c} ( \epsilon)$ within the large-$N$ approximation,
with $N=2$.}
\end{figure}

\section{Self-Consistent Harmonic Approximation}\label{SSCHA}

\begin{figure} 
\includegraphics[width=9cm]{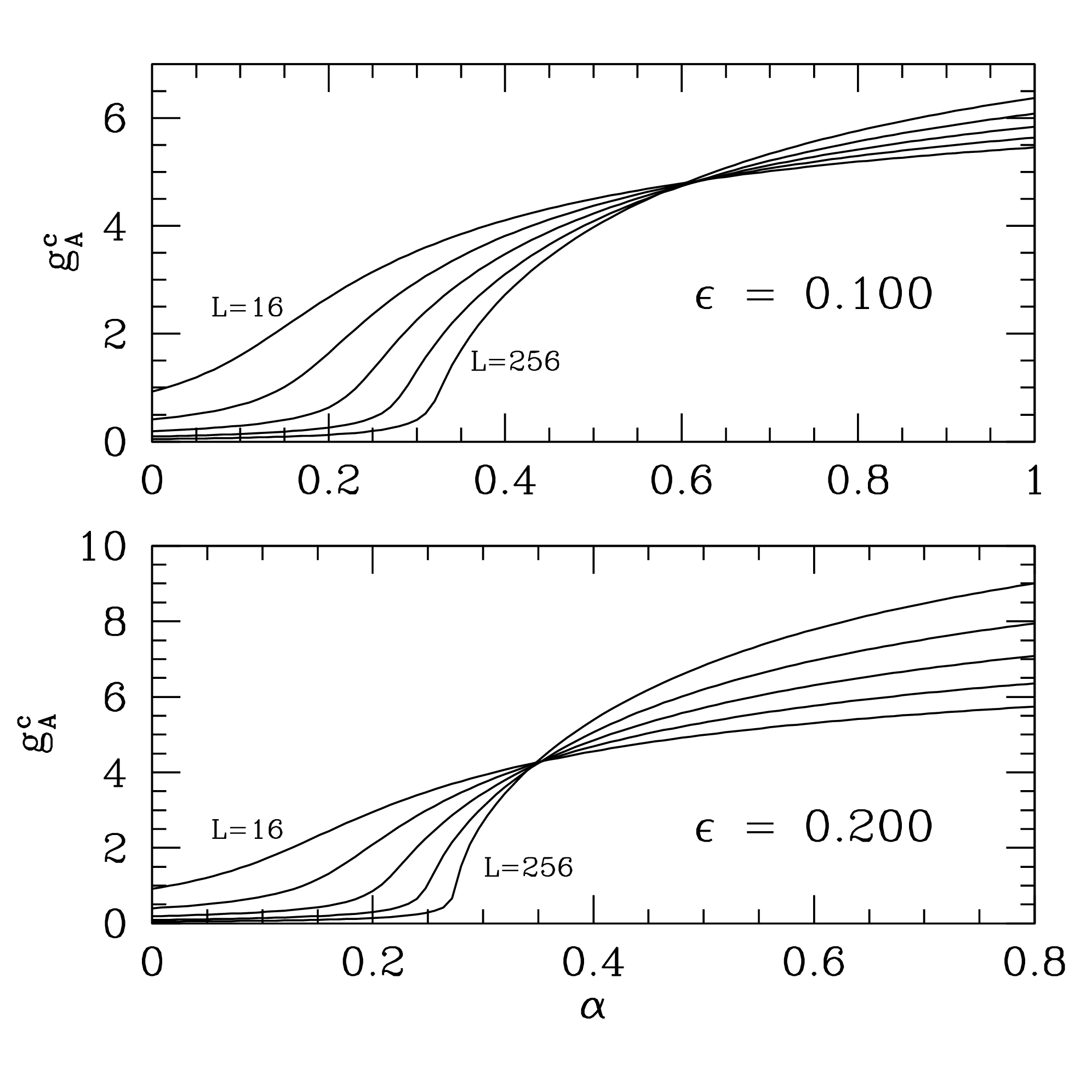}
\caption
{\label{scha1}  Results of the self-consistent harmonic approximation for $ \epsilon=0.10$ and
$ \epsilon=0.20$, with dimensionless inverse temperature $L=16,32,64,128,256$.  The crossing
of the curves at a single point indicates a second order phase transition.}
\end{figure}

\begin{figure} 
\includegraphics[width=9cm]{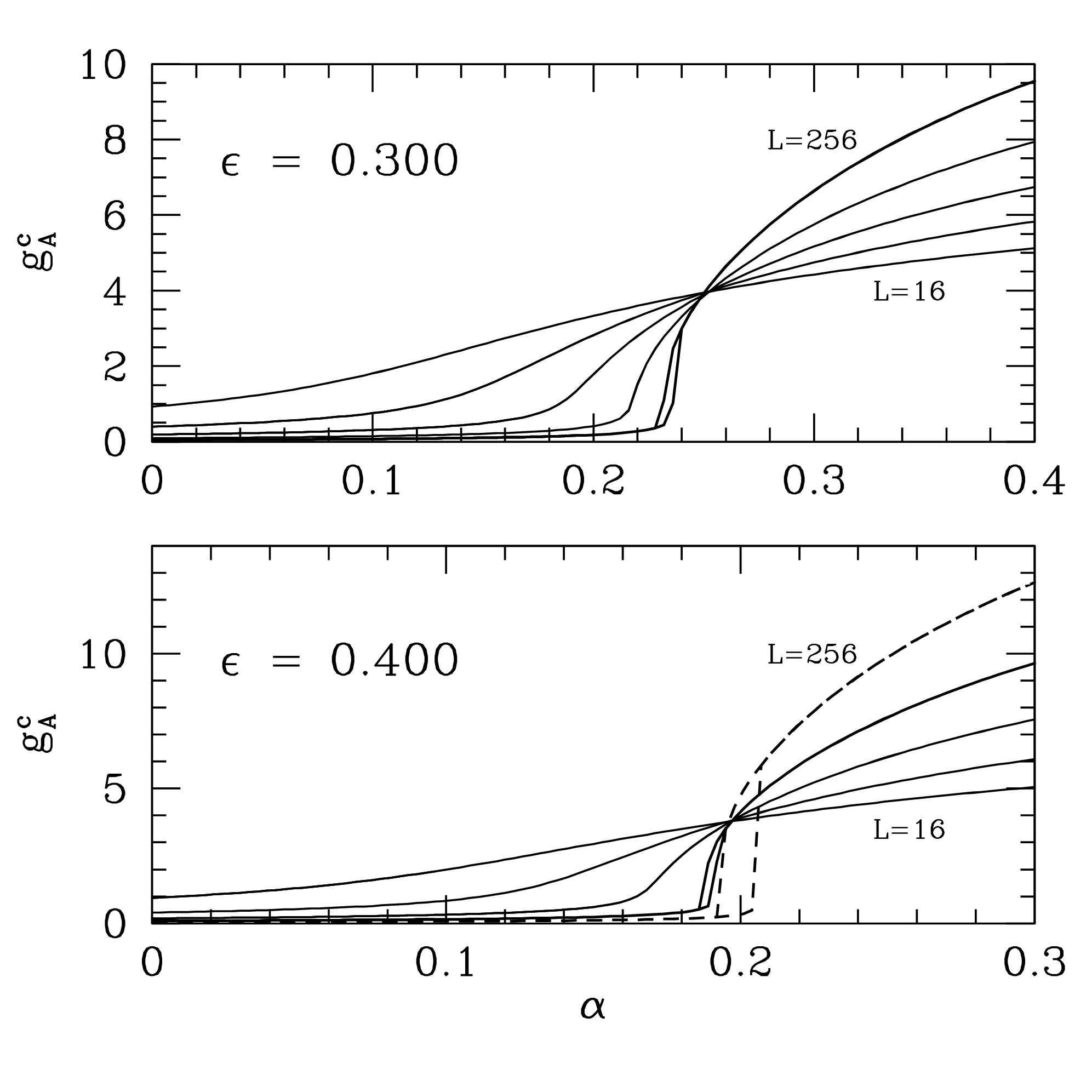}
\caption
{\label{scha2}  Results of the self-consistent harmonic approximation for $ \epsilon=0.30$ and
$ \epsilon=0.40$, with dimensionless inverse temperature $L=16,32,64,128,256$.   For high
temperatures, the curves seem to cross at a single point.  However, a spurious first order
transition intervenes at low $T$ (large $L$).}
\end{figure}

The fully nonlinear theory with action
\begin{eqnarray}
&&S[\phi(s)]= \int\limits_0^L\!\! ds\,{1\over 4}\left({ {\partial}\phi\over {\partial} s}\right)^2\\
&&\quad+\,\alpha \int\limits_0^L\!\! ds \int\limits_0^L\!\! ds'\,K(s-s')
\big[1-\cos\big(\phi(s)-\phi(s')\big)\big]\nonumber
\end{eqnarray}
is replaced with a trial Gaussian theory described by the quadratic action
\begin{eqnarray}
S_0[\phi(s)]&=& \frac{1}{2} \int\limits_0^L\!\! ds \int\limits_0^L\!\! ds'\,V(s-s')\,
\big[\phi(s)-\phi(s')\big]^2\nonumber\\
&=& \frac{1}{2}\sum_{\omega_n} {\hat G}^{-1}(\omega_n)\,| {\hat\phi} (\omega_n)|^2\ ,
\end{eqnarray}
where
\begin{equation}
 {\hat G} (\omega_n)\equiv{1\over 2\big[ {\hat V} (0)- {\hat V} (\omega_n)\big]}\ .
\end{equation}
$V(s)$, or equivalently $ {\hat G} (\omega_n)$, is treated variationally,
so we extremize by setting
\begin{equation}
{ {\partial}\over {\partial} {\hat G} (\omega_n)}\left\{F_0 + {1\over L}\,
\langle S - S_0\rangle ^{\vphantom{\dagger}}_0\right\}
\end{equation}
where $F_0=-L^{-1} \mathop{\rm Tr}_\phi \exp(-S_0[\phi])$.  This leads to the following
self-consistent equation:
\begin{equation}
{1\over {\hat G} (\omega_n)}= \frac{1}{2}\,\omega_n^2 + 2\alpha \int\limits_0^L\!\! ds\,
\big(1-\cos(\omega_n s)\big)\,K(s)\,\Gamma(s)
\end{equation}
where
\begin{eqnarray}
\Gamma(s)&\equiv&\exp\left(- \frac{1}{2}\Big\langle\left[\phi(s)-\phi(0)\right]^2
\Big\rangle ^{\vphantom{\dagger}}_0\right)\nonumber\\
&=&\exp\left(-{2\over L}\sum_{n=1}^\infty\big(1-\cos(\omega_n s)\big)
\, {\hat G} (\omega_n)\right)\ .
\end{eqnarray}
We iterate these equations to self-consistency,

The dimensionless conductance $g_{\scriptscriptstyle{\rm A}}(\alpha,L)$ is plotted for $ \epsilon=0.1$ and $0.2$
in fig. \ref{scha1}.  As the dimensionless inverse temperature $L$ is increased from $L=16$
to $L=256$, the curves apparently cross at a critical point, $\alpha_ {\rm c} ( \epsilon)$.  For
$\alpha>\alpha_ {\rm c}$, the conductance increases as the temperature is lowered, indicating
a conducting phase.  For $\alpha < \alpha_ {\rm c}$ the conductance vanishes as the temperature
is lowered,  {\it i.e.\/}\ the Coulomb gap survives.  For larger values of $ \epsilon$, however, a spurious
first order transition preempts this critical behavior, as shown in fig. \ref{scha2}.  The solution
to the SCHA is hysteretic, and discontinuous, provided $L$ is large enough.

\section{Coulomb Gas Representation and Monte Carlo}\label{SMC}
Including the offset charge, the Euclidean action for our system is
$S=S_0+S_{\rm int}+S_{\rm top}$, where
\begin{eqnarray}
S_0[\phi]&=&{1\over 4} \int\limits_0^L\!\! ds\,\left({ {\partial}\phi\over {\partial} s}\right)^2\\
S_{\rm int}[\phi]&=&-\alpha\! \int\limits_0^L\!\! ds\! \int\limits_0^L\!\! ds'\,K(s-s')\,\cos
\left[\phi(s)-\phi(s')\right]\\
S_{\rm top}&=&-i\nu\left[ \phi(L)-\phi(0)\right]\equiv -2\pi i\nu W[\phi]\ ,
\end{eqnarray}
where $W[\phi]=[\phi(L)-\phi(0)]/2\pi$ is the winding number of the 
phase field.  In the phase representation, the topological term represents
a purely imaginary contribution to the action.  However, in the Coulomb gas
representation, the action remains purely real.

We have discarded a formally divergent constant from our action,
\begin{equation}
\Delta S=L \int\limits_0^L\!\! ds\,K(s)\ ,
\end{equation}
which may be rendered finite through an ultraviolet regularization of
$K(s)$, {\it viz.}
\begin{equation}
K_b(s)=\left\{{\pi\over L}\csc\left({\pi |s|\over L}\right)\cdot
\left[1-e^{-L\sin(\pi |s|/L)/\pi b}\right]\right\}^{2- \epsilon}\ ,
\label{bcut}
\end{equation}
so that $K_b(0)=b^{-(1- \epsilon)}$. 

The partition function for our problem is given by
\begin{eqnarray}
\Xi(\nu)&=&\Xi_0\Big\langle e^{i\nu[\phi(L)-\phi(0)]}\,\sum_{N=0}^\infty
{1\over N!}\,(-S_{\rm int})^N\Big\rangle ^{\vphantom{\dagger}}_0\\
&=&\Xi_0\sum_{N=0}^\infty{\alpha^N\over N!} \int\limits_0^L\!\! ds^+_1\cdots \int\limits_0^L\!\! ds^-_N
\prod_{j=1}^N K(s_j^+-s_j^-)\nonumber\\
&&\quad\times\Big\langle e^{i\nu\phi(L)} e^{-i\nu\phi(0)}\,
e^{i\sum_{k=1}^N\left[\phi(s^+_k)-\phi(s^-_k)\right]}\Big\rangle ^{\vphantom{\dagger}}_0
\nonumber\\
\end{eqnarray}
where the average is with respect to the bare action $S_0$.  This allows us to
transform the complex action (due to the topological term) in the phase representation
to a purely real action in a dipole gas representation\cite{gs85}.  The coordinates
$\{s^\pm_k\}$ are interpreted as locations of positive and negative charges.
Each factor of $(-S_{\rm int})$ introduces another pair of such charges,
 {\it i.e.\/}\ a dipole.  We now write the field $\phi(s)$ as a sum of winding plus a
periodic pieces:
\begin{eqnarray}
\phi(s)&=&{2\pi W s\over L}+\eta+ {\bf\varphi} (s)\\
 {\bf\varphi} (s)&=&{1\over\sqrt{L}} \mathop{{\sum}^\prime}_{\omega_n}
 {\hat {\bf\varphi}} (\omega_n)\,e^{-i\omega_n s}\ ,
\end{eqnarray}
where $\eta$ is a constant and $W$ is the winding number.  The prime on
the sum denotes exclusion of the $n=0$ term, which is accounted for by
$\eta$.  The bare action is then
\begin{equation}
S_0={\pi^2 W^2\over L} + \frac{1}{4} \mathop{{\sum}^\prime}_{\omega_n} \omega_n^2\,
| {\hat {\bf\varphi}} (\omega_n)|^2
\end{equation}
Thus, $\langle | {\hat {\bf\varphi}} (\omega_n)|^2\rangle ^{\vphantom{\dagger}}_0=2/\omega_n^2$, and
\begin{eqnarray}
C(s)&\equiv& \frac{1}{2}\Big\langle \left[  {\bf\varphi} (s)- {\bf\varphi} (0)\right]^2\Big\rangle ^{\vphantom{\dagger}}_0
\nonumber\\
&=&{L\over\pi^2}\sum_{n=1}^\infty {1-\cos(2\pi ns/L)\over n^2}\nonumber\\
&=&\left[ s \left(1-{s\over L}\right)\right]_{\rm per}\ ,
\end{eqnarray}
where the subscript indicates that the expression is to be periodically
extended from its value on the interval $s\in [0,L]$.
Summing over the winding number $W$ and
averaging over $\phi(s)$, we obtain the $N$-dipole pair Boltzmann weight,
\begin{eqnarray}
&& \varrho ^{\vphantom{\dagger}}_N(s^+_1,\ldots,s^-_N)={\alpha^N\over N!}\,
\prod_{j=1}^N K(s^+_j-s^-_j)\times\\
&&\quad\exp\left[ \frac{1}{2}\sum_{j,j'\atop\sigma,\sigma'} \sigma\sigma'\,
C(s_j^\sigma-s_{j'}^{\sigma'})\right]
\cdot{ \vartheta_3(\pi P/L+\pi\nu\,|\,i\pi/L)\over \vartheta_3(0\,|\,i\pi/L)}\ ,\nonumber
\end{eqnarray}
where $P$ is the total dipole moment,
\begin{equation}
P=\sum_{j=1}^N(s^+_j-s^-_j)\ ,
\end{equation}
and $ \vartheta_3(z\,|\,\tau)$ is the Jacobi theta-function\cite{wwcma}.

\begin{figure}[t]
\includegraphics[width=9cm]{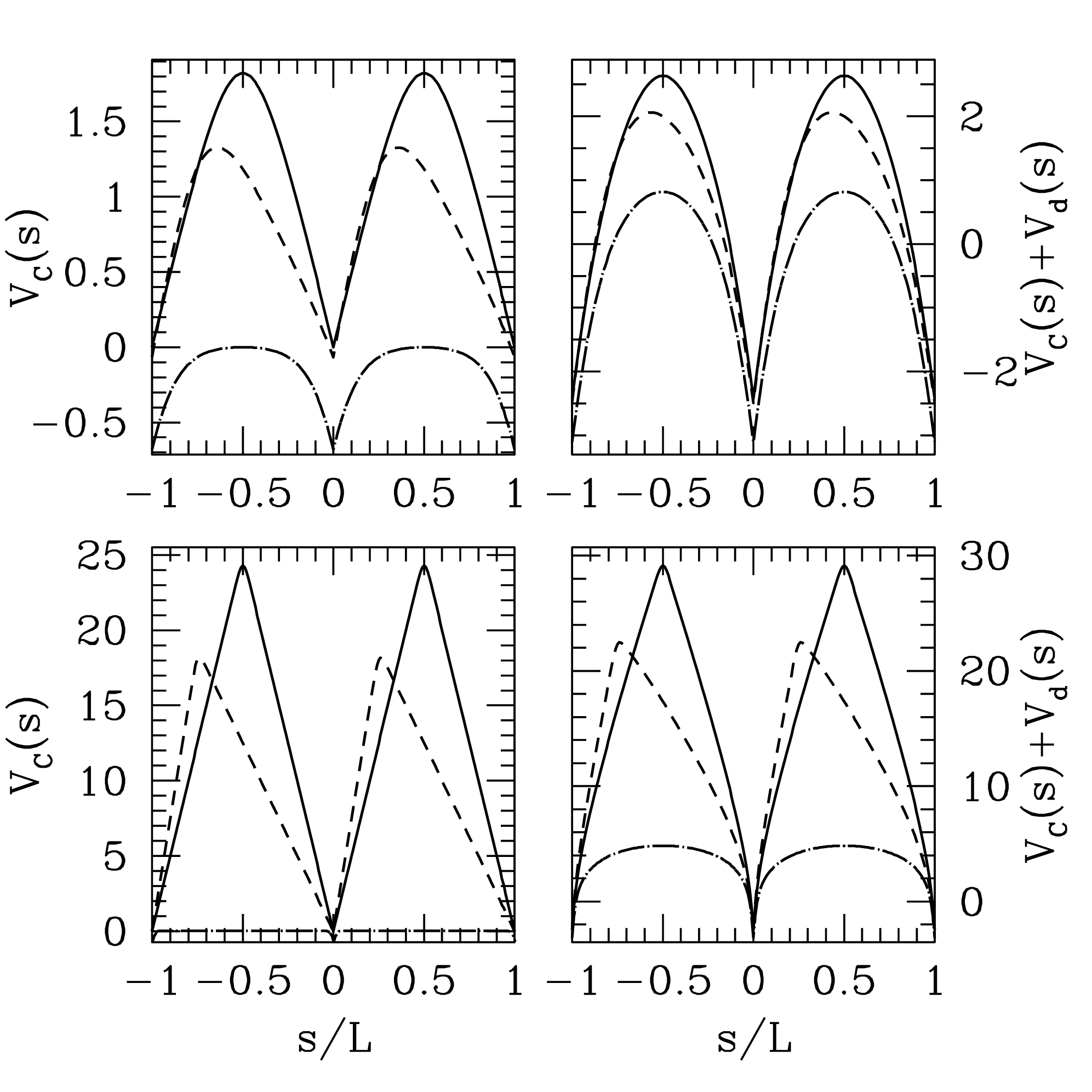}
\caption
{\label{pair} Single pair potentials $V_ {\rm C} (s)$ and
$V_ {\rm C} (s)+V_ {\rm d} (s)$ for $\nu=0$ (solid), $\nu=0.25$ (dashed), and 
$\nu=0.5$ (dot-dashed).  Upper panels have $L=5$; lower panels have $L=25$.
In all cases $b=0.2$.}
\end{figure}

\begin{figure} [t]
\includegraphics[width=9cm]{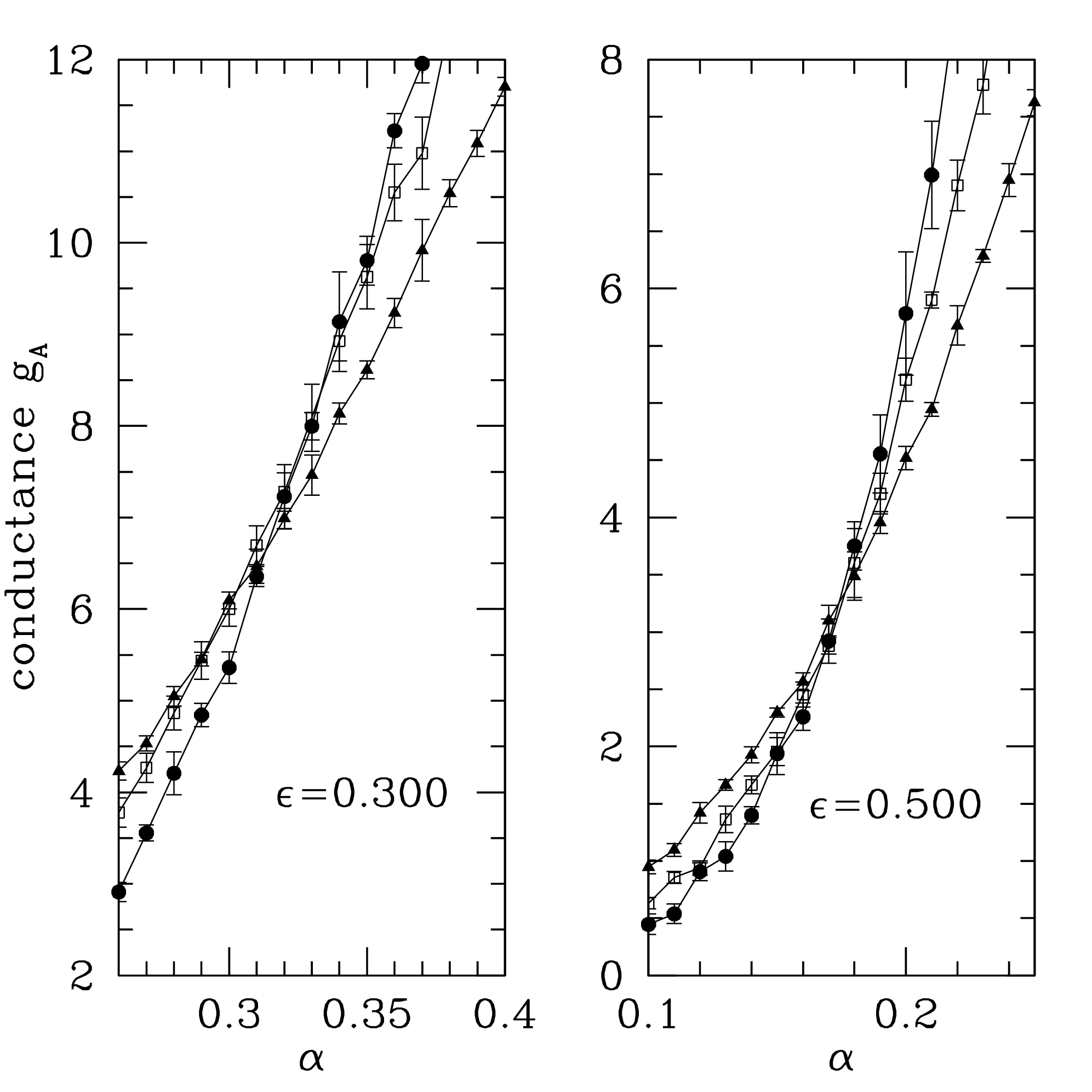}
\caption
{\label{mcraw} Coulomb gas Monte Carlo results for conductance $g_{\scriptscriptstyle{\rm A}}(\alpha)$
at $L=10$ (filled triangles), $L=20$ (open squares), and $L=40$ (filled circles) for $ \epsilon=0.30$
(left panel) and $ \epsilon=0.50$ (right panel).  A critical point marks the boundary between a
high $\alpha$ conducting phase and a low $\alpha$ insulating (Coulomb blockade) phase.}
\end{figure}

We now use the $ \vartheta$-function identity,
\begin{equation}
 \vartheta_3(z\,|\,\tau)=\left({i\over\tau}\right)^{1/2}\,e^{-iz^2/\pi\tau}
 \vartheta_3\Big(-{z\over\tau}\,\Big|\,-{1\over\tau}\Big)
\end{equation}
to obtain
\begin{eqnarray}
 \varrho ^{\vphantom{\dagger}}_N(s^+_1,\ldots,s^-_N)&=&{\alpha^N\over N!}\,
\prod_{j=1}^N K(s^+_j-s^-_j)\cdot{H_L(P+\nu L)\over H_L(0)}\nonumber\\
&&\qquad \times e^{-U}\,e^{-2\nu P}\,e^{-L\nu^2}\ ,
\label{bweight}
\end{eqnarray}
where $U$ is given by
\begin{equation}
U=- \frac{1}{2}\sum_{j,j'\atop\sigma\sigma'} \sigma\sigma'\,\big|s^\sigma_j
-s^{\sigma'}_{j'}\big|
\end{equation}
and
\begin{eqnarray}
H_L(y)&=& \vartheta_3(iy\,|\,iL/\pi)\nonumber\\
&=&\sum_{m=-\infty}^\infty e^{-m^2 L}\,e^{-2my}\ ,
\end{eqnarray}
and where the location of each charge is restricted to the interval
$s^\sigma_j\in [0,L]$, so that $P\in [-NL,NL]$.

Interpreting (\ref{bweight}) for $\nu=0$, we have that the system consists of
a gas of $N$ dipole pairs.  Each of the $2N$ charges interacts with every
other charge according to a one-dimensional Coulomb potential
$-\sigma\sigma'|s-s'|$.  In addition, each charge is logarithmically attracted
to its unique mate - the other member of its dipole pair - via a potential
$V ^{\vphantom{\dagger}}_ {\rm d} (s)=-\ln K(s)\simeq (2- \epsilon)\ln |s|$.  For $\nu\ne 0$ there
is an additional electric field of strength $2\nu$ present.  $\alpha$ is
the dipole fugacity; $\alpha\to 0$ will suppress the appearance of dipoles.
Hence, in the absence of any short-time cutoff $b$, short-distance dipole pairs
will proliferate without limit.  A simple estimation of the dipole density, ignoring
the Coulomb interactions, is $n_{\rm dip}\sim b^{-(1- \epsilon)}/(1- \epsilon)$.
In our Monte Carlo simulations we have used $b=1$; the essential
physics is rather weakly dependent on $b$, though.
Grand canonical averages are to be computed in the usual way,  {\it i.e.\/}\ 
$\langle A\rangle =  \mathop{\rm Tr} ( \varrho A)/ \mathop{\rm Tr} \varrho$, where the trace entails a sum over
all possible numbers $N$ of dipole pairs and integration over their $2N$ charge
coordinates.

In figure \ref{pair} we plot the Coulomb energy for a single dipole pair,
\begin{equation}
V ^{\vphantom{\dagger}}_ {\rm C} (s)=|s|+\ln {H_L(s+\nu L)\over H_L(0)} + 2\nu s
\end{equation}
as well as the total dipole energy $V ^{\vphantom{\dagger}}_ {\rm C} (s)+V ^{\vphantom{\dagger}}_ {\rm d} (s)$ for $\nu=0$,
$\nu=\frac{1}{4}$, and $\nu= \frac{1}{2}$ for $L=5$ and $L=25$.  Note how the
interaction becomes asymmetric (yet still properly periodic) when $\nu$ is
neither integer nor half odd integer.  Furthermore, the Coulomb interaction
is effectively cancelled when $\nu= \frac{1}{2}$.

The XY phase correlation function,
\begin{equation}
G(s)=\langle e^{i\phi(s)}\,e^{-i\phi(0)}\rangle\ ,
\end{equation}
is simply related to the dipole separation correlation function,
\begin{equation}
h(s)=\Big\langle\sum_{i=1}^N\delta(s^+_i-s^-_i-s)\Big\rangle .
\end{equation}
To see this, define the quantity
\begin{eqnarray}
Q ^{\vphantom{\dagger}}_N(\tau^+,\tau^-)&=&(N+1) \int\limits_0^L\!\! ds^+_1 \int\limits_0^L\!\! ds^-_1\cdots \int\limits_0^L\!\! ds^+_N \int\limits_0^L\!\! ds^-_N
\nonumber\\
&& \varrho ^{\vphantom{\dagger}}_{N+1}(s^+_1,s^-_1,\ldots,s^+_N,s^-_N,\tau^+,\tau^-)\ .
\end{eqnarray}
One then has
\begin{eqnarray}
\alpha\,K(s)\,G(s)&=&{1\over L\Xi}\sum_{N=0}^\infty \int\limits_0^L\!\! d\tau\,
Q ^{\vphantom{\dagger}}_N(\tau+s,\tau)\nonumber\\
&=&h(s)\ .
\end{eqnarray}
What is computed in the Coulomb gas Monte Carlo calculation is
the dipole separation correlation function $h(s)$, from which the
XY phase correlator is obtained via $G(s)=h(s)/\alpha K(s)$.  The 
conductance is then obtained using (\ref{gaeqn}).

\begin{figure} [t]
\includegraphics[width=9cm]{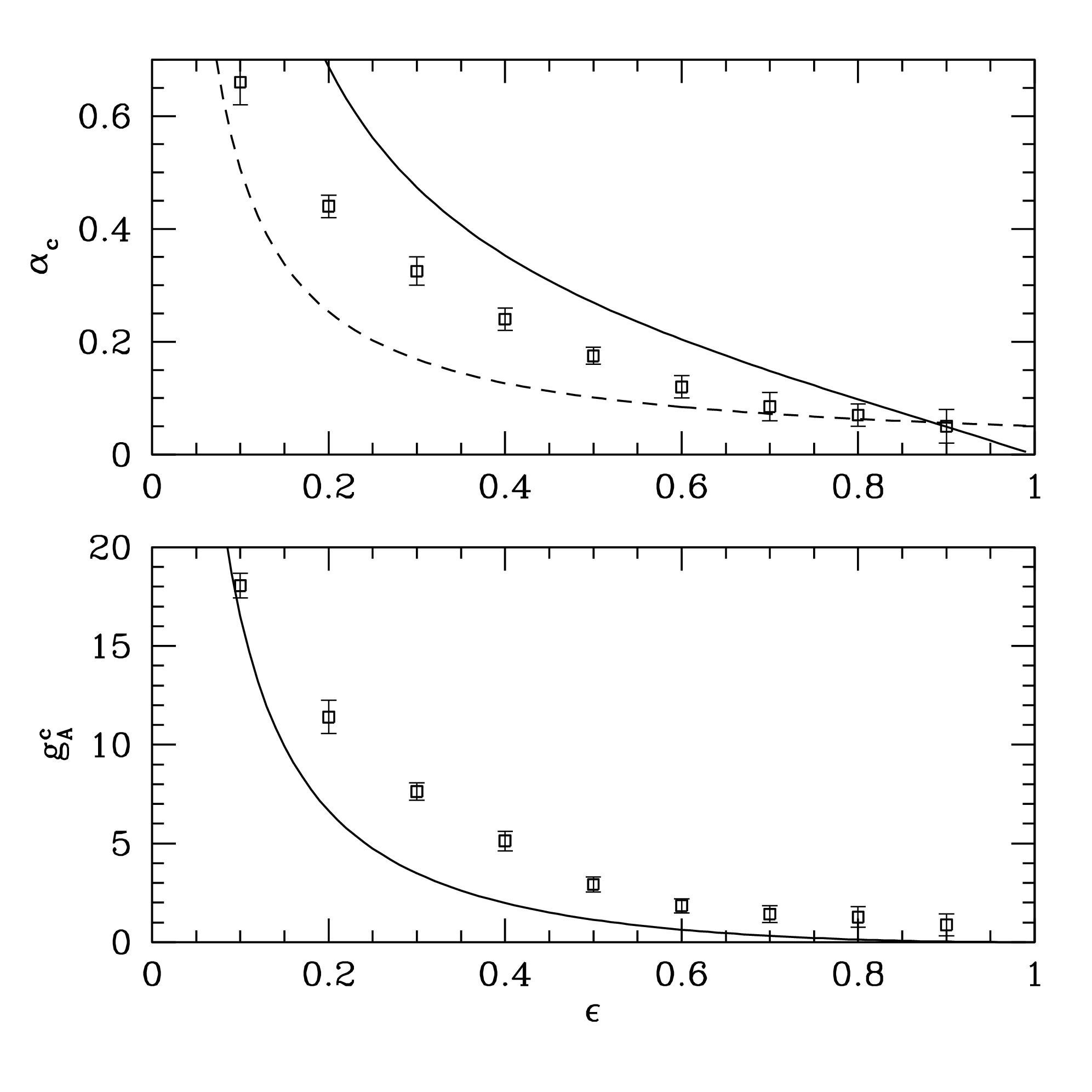}
\caption
{\label{dcrit} Critical coupling $\alpha_ {\rm c} ( \epsilon)$ (top) and critical
conductance $g_{\scriptscriptstyle{\rm A}}^{\rm c} ( \epsilon)$ (bottom), comparing data
from Monte Carlo (points) and large-$N$ (smooth curve) calculations.
The dashed line in the top figure is the Kosterlitz RG result $ \alpha_ {\rm c}^{\rm RG}=1/2\pi^2 \epsilon$.}
\end{figure}

The renormalized charging energy, $E_ {\rm c}^*$, is given by
\begin{eqnarray}
{E_ {\rm c}^*\over E_ {\rm c}}&=&{1\over 2}{ {\partial}^2\! F\over {\partial}\nu^2}
\Big|_{\nu=0}\\
&=&1-{L\over 2}\, \Big\langle {H^{\prime\prime}_L(P)\over H_L(P)}\Big\rangle
+2\,\Big\langle P\,{H^\prime_L(P)\over H_L(P)}\Big\rangle
-{2\over L}\,\langle P^2\rangle\ .\nonumber
\end{eqnarray}
Note that when the dipole fugacity $\alpha$ vanishes, there are no pairs
at all and one obtains
\begin{equation}
{E_ {\rm c}^*\over E_ {\rm c}}\Big|_{\alpha=0}=1-{L\over 2}\,{H^{\prime\prime}_L(0)\over H_L(0)}\ ,
\end{equation}
which also follows from an analysis of the noninteracting Hamiltonian
$ {\cal H}_{\alpha=0}=E_ {\rm c}\,({\hat n}+\nu)^2$.  In this case the renormalized
charging energy $E_ {\rm c}^*$ interpolates between its low temperature value
of $E_ {\rm c}^*=E_ {\rm c}$ and its high temperature limit of $E_ {\rm c}^*=0$.

\subsection{Results}
Typical raw data for $g_{\scriptscriptstyle{\rm A}} (\alpha,L)$ are shown in fig. \ref{mcraw}.  Two phases are
identified for every positive value of $ \epsilon$: a small $\alpha$ insulating phase in which
the Coulomb gap persists and $g_{\scriptscriptstyle{\rm A}}$ decreases as $T\to 0$ ($L\to\infty$), and a
large $\alpha$ conducting phase in which $g_{\scriptscriptstyle{\rm A}}$ diverges as $T\to 0$, indicating
a nonlinear $I-V$ relation at $T=0$.  The boundary between these phases is marked
by a critical value $\alpha_ {\rm c} ( \epsilon)$, at which there is a quantum phase transition.
At $\alpha=\alpha_ {\rm c}$, the conductance becomes temperature-independent (provided
$T$ is low enough that the system is in the scaling regime).    Monte Carlo results for $\alpha_ {\rm c} ( \epsilon)$ and $g_ {\rm A}^ {\rm c} ( \epsilon)$ are presented in fig. \ref{dcrit},
along with comparisons to large-$N$ results.  

\begin{figure} [t]
\includegraphics[width=9cm]{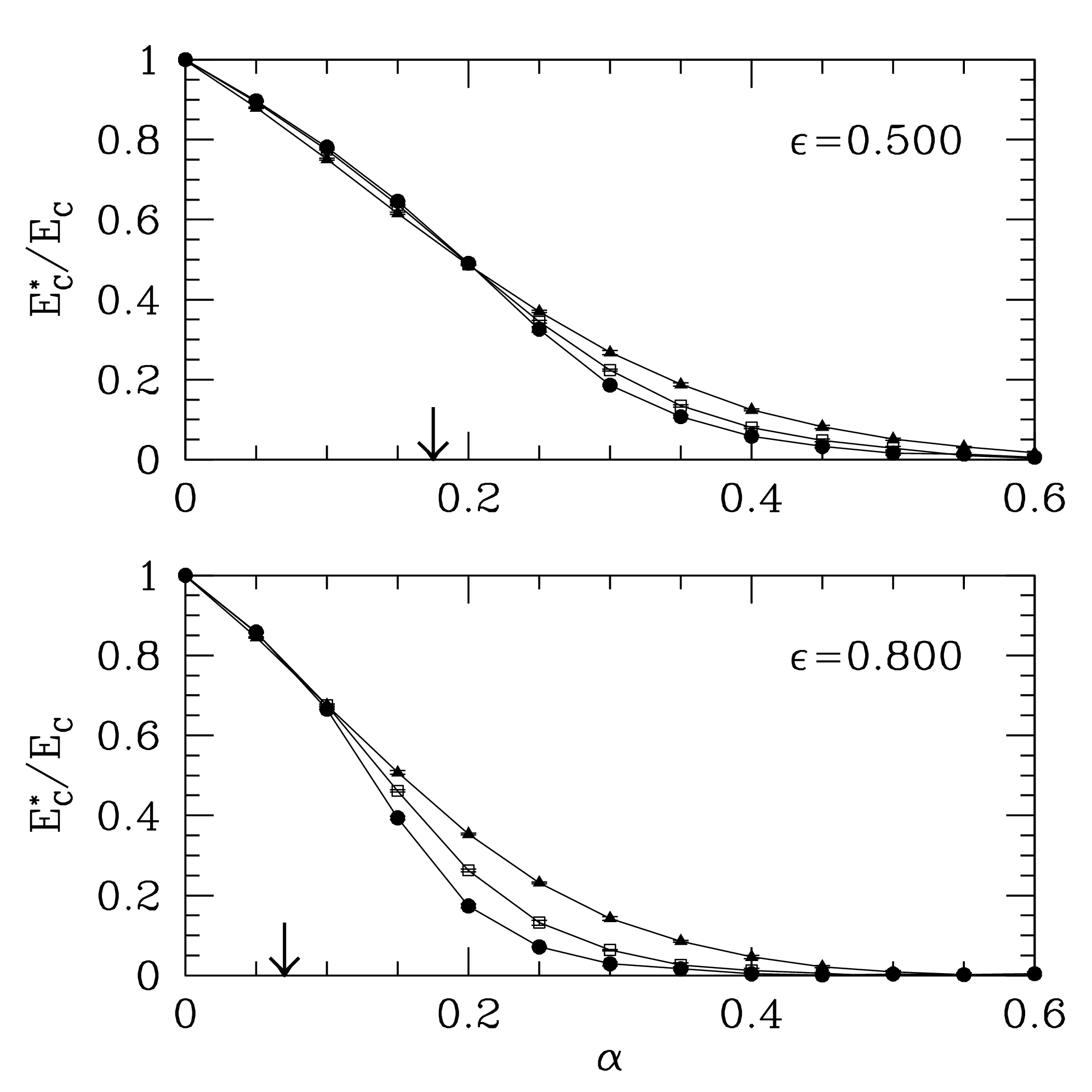}
\caption
{\label{echg} Dimensionless renormalized charging energy $E_ {\rm c}^*/E_ {\rm c}$ {\it versus\/}
$\alpha$ at $L=10$ (filled triangles), $L=20$ (open squares), and $L=40$ (filled circles) for $ \epsilon=0.50$ (top panel) and $ \epsilon=0.80$ (bottom panel).  Arrows indicate the location of
the phase transition as obtained from crossing of conductance curves.}
\end{figure}

\begin{figure} [t]
\includegraphics[width=9cm]{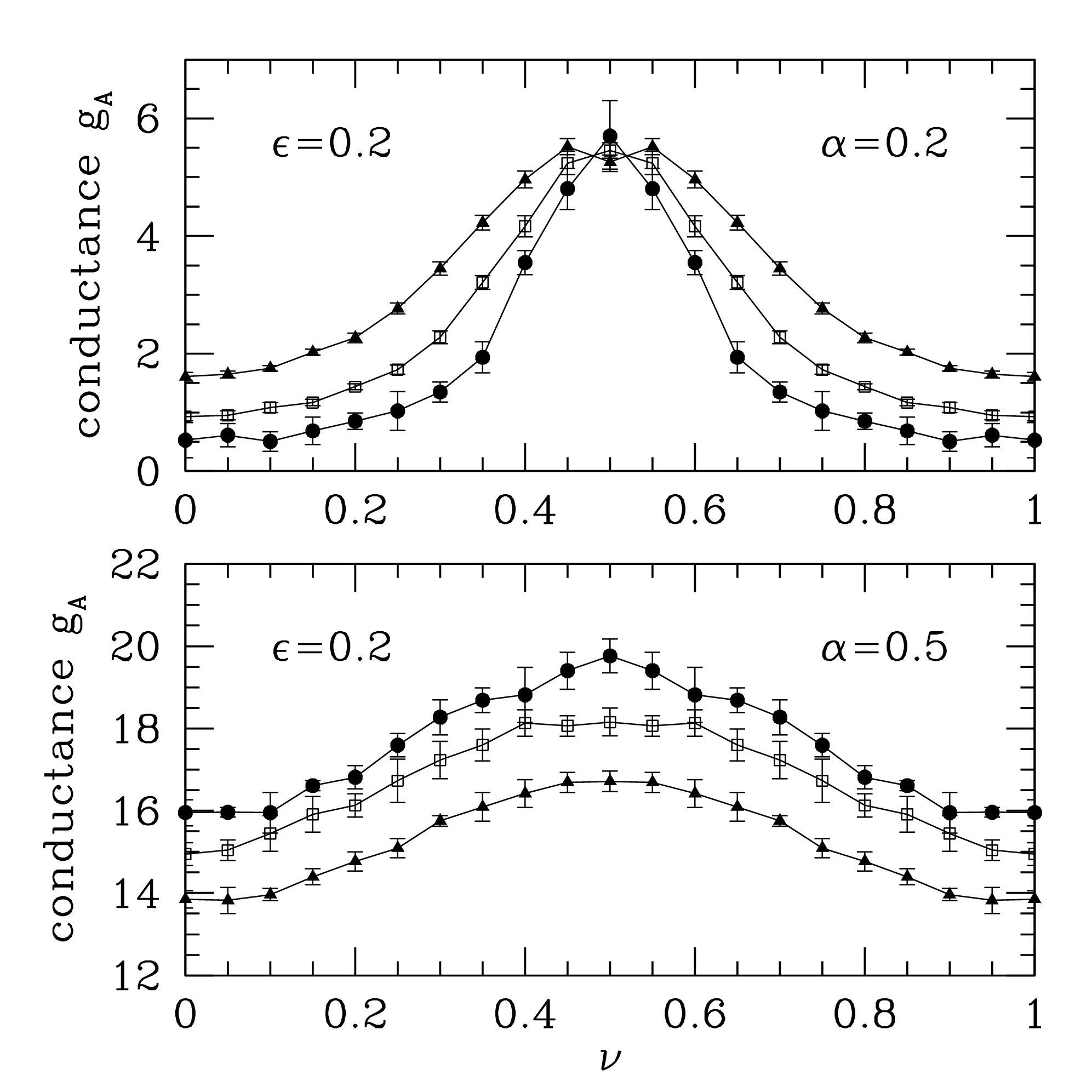}
\caption
{\label{q2} Conductance $g_{\scriptscriptstyle{\rm A}}$ {\it versus\/} charge offset $\nu$ at $ \epsilon=0.2$ for
$\alpha=0.2$ (top panel) and $\alpha=0.5$ (bottom panel) at dimensionless inverse 
temperatures $L=10$ (filled triangles), $L=20$ (open squares), and $L=40$ (filled circles).
The critical coupling for $\nu=0$ is $\alpha_ {\rm c}\simeq 0.44.$ }
\end{figure}

Our results for $\alpha_ {\rm c} ( \epsilon)$ differ significantly from those of Bascones {\it et al.}\cite{bas00}, who obtained $\alpha_ {\rm c}$ through analysis of the renormalized charging energy, assuming
$E_ {\rm c}^*(\alpha)\sim (\alpha_ {\rm c}-\alpha)^{1/ \epsilon}$.  (Our values for $\alpha_ {\rm c} ( \epsilon)$ are
approximately four times smaller throughout the range $0 <  \epsilon \le  \frac{1}{2}$.)  In fact, we find that our
raw data for $E_ {\rm c}^*/E_ {\rm c}$ {\it versus\/} $\alpha$ are in good agreement with those of
Bascones {\it et al.}  \cite{agree}.  This agreement is noteworthy since their Monte Carlo was carried out in the phase representation whereas our is in the Coulomb gas representation.  However, it is very difficult to reliably extract $\alpha_ {\rm c}$ from the charging energy data, as our results shown in  fig. \ref{echg} show.  At a temperature corresponding to $L=50$ 
(ref. \onlinecite{bas00}) or $L=40$ (this work), there is no detectable signature of the phase transition at $\alpha=\alpha_ {\rm c}$. Similar behavior is found in the large-$N$ results of fig. \ref{lnc}.  In the large-$N$ theory, $\lambda(\alpha)$ plays the role of an energy gap, similar to $E_ {\rm c}^*$.  As is evident from fig. \ref{lnc}, even at relatively low temperatures of
$1/L\approx 1/32$, extrapolation of $\alpha_ {\rm c}$ based on the zero-temperature critical
( {\it i.e.\/}\ power law) behavior is problematic.  Indeed, for fixed $1/L$, varying $\alpha$ takes
the system through renormalized classical, quantum critical, and quantum disordered
regimes\cite{sac99}, and the single parameter $L=\infty$ behavior of $\lambda(\alpha)$ is insufficient to extract $\alpha_ {\rm c}$\cite{spurious}.

\begin{figure} [t]
\includegraphics[width=9cm]{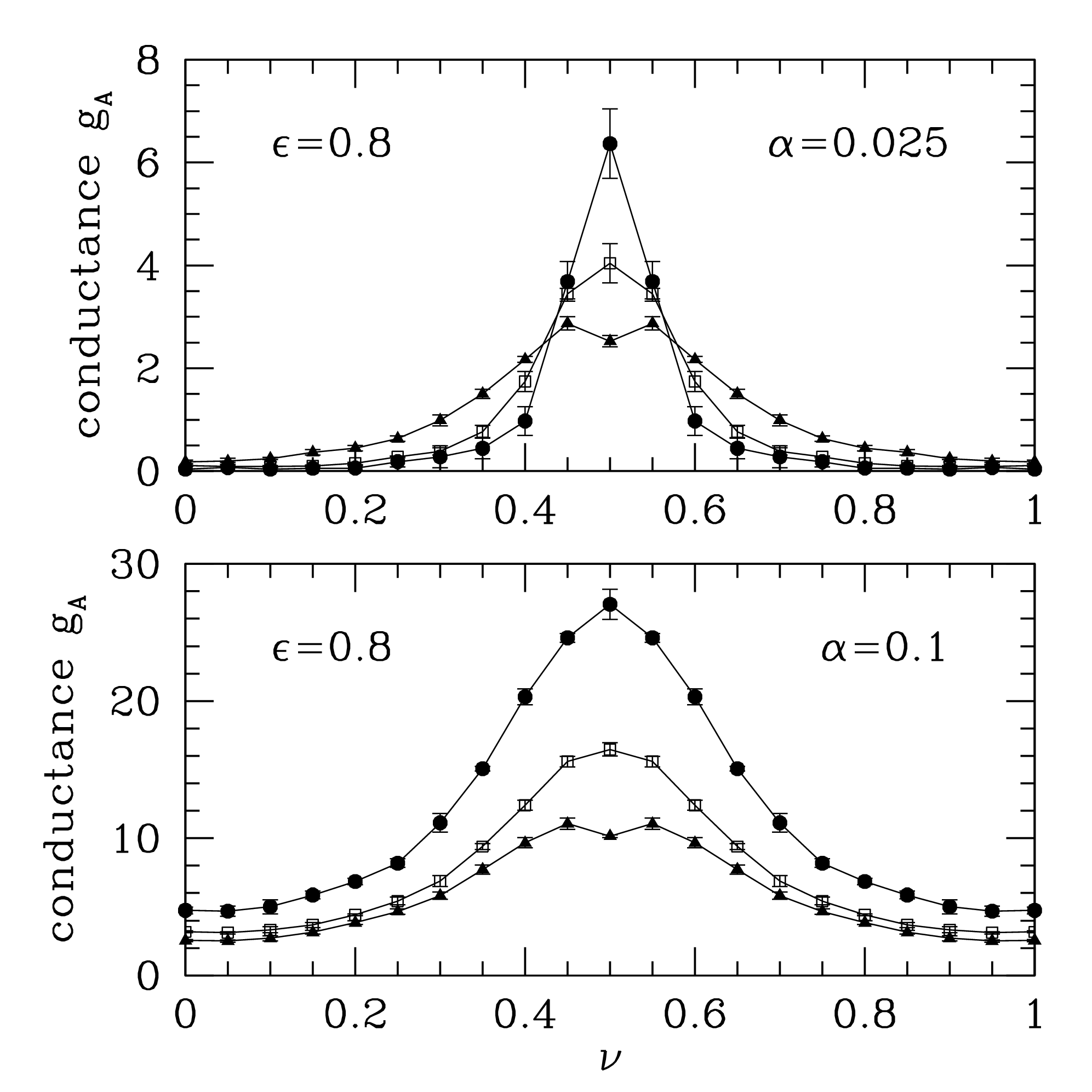}
\caption
{\label{q8} Conductance $g_{\scriptscriptstyle{\rm A}}$ {\it versus\/} charge offset $\nu$ at $ \epsilon=0.8$ for
$\alpha=0.025$ (top panel) and $\alpha=0.10$ (bottom panel) at dimensionless inverse 
temperatures $L=10$ (filled triangles), $L=20$ (open squares), and $L=40$ (filled circles).
The critical coupling for $\nu=0$ is $\alpha_ {\rm c}\simeq 0.07.$ }
\end{figure}

\begin{figure} [t]
\includegraphics[width=9cm]{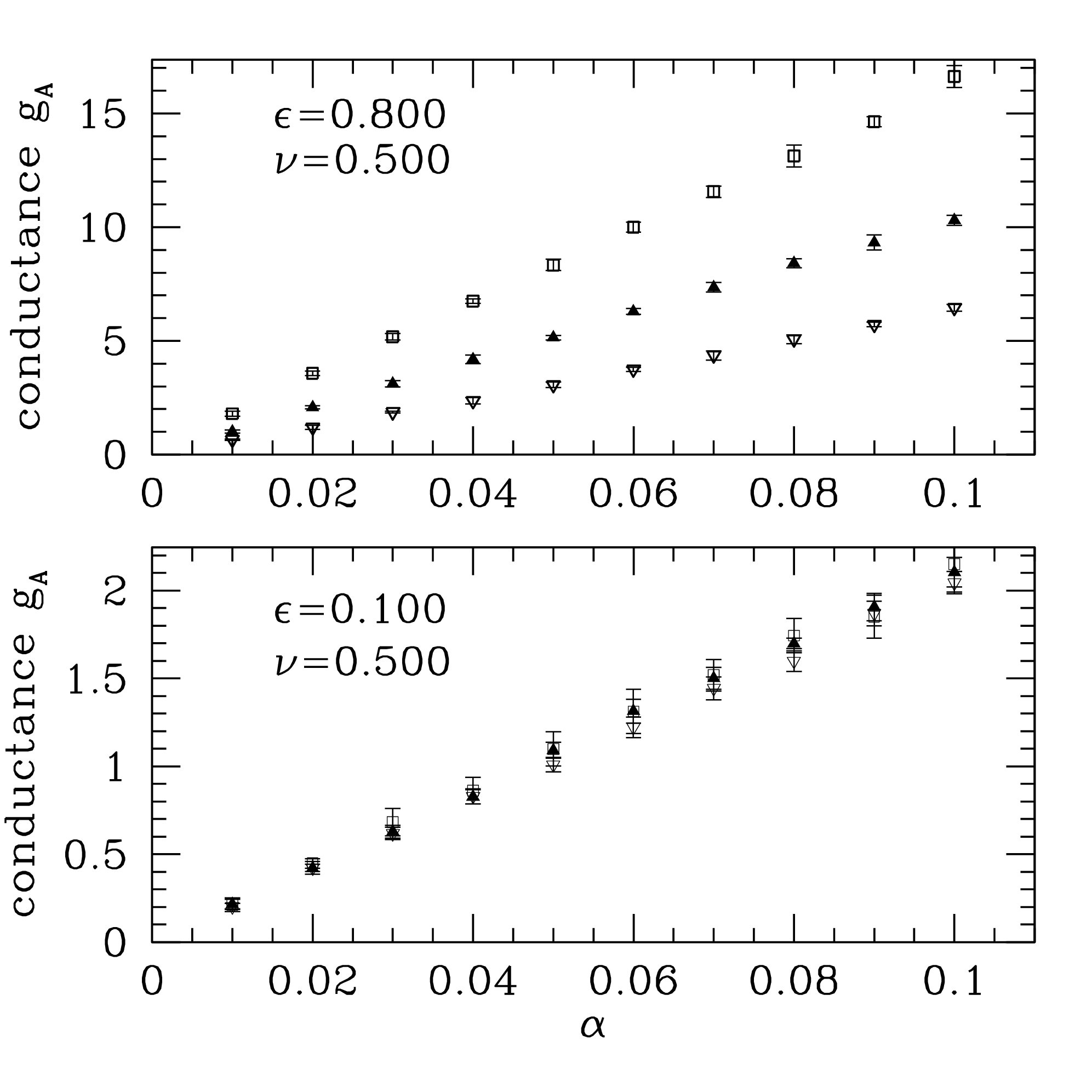}
\caption
{\label{nuhalf} Conductance {\it versus\/} coupling $\alpha$ at charge offset $\nu= \frac{1}{2}$
for $ \epsilon=0.80$ (top) and $ \epsilon=0.1$ (bottom), for $L=5$ (open down-pointing triangles),
$L=10$ (filled triangles), and $L=20$ (open squares).  For reference, at  $\nu=0$, 
$\alpha_ {\rm c} ( \epsilon=0.8)\simeq 0.07$ and $ \alpha_ {\rm c} ( \epsilon=0.1)\simeq 0.66$.}
\end{figure}

Finally, we plot conductance {\it versus\/} charge offset $\nu$ for $ \epsilon=0.2$ and $ \epsilon=0.8$\ in
figs. \ref{q2} and \ref{q8}, respectively.  Two values of $\alpha$ on either side of $\alpha_ {\rm c}$
are chosen, corresponding to opposite temperature dependences at $\nu=0$.   We find a
curious double peak structure in the vicinity of $\nu= \frac{1}{2}$ at higher temperatures, but which
disappears as $T\to 0$.  For $\nu= \frac{1}{2}$ and $ \epsilon=0.8$
the conducting state prevails even at small values of $\alpha$.  This feature is emphasized
in fig. \ref{nuhalf}, where $g_{\scriptscriptstyle{\rm A}}(\alpha,\nu= \frac{1}{2})$ is contrasted for $ \epsilon=0.8$ and
$ \epsilon=0.1$.  For $ \epsilon=0.8$, a conducting state is observed down to values of $\alpha$
as small as $0.01$, well below the $\nu=0$ critical value of $\alpha_ {\rm c}\simeq 0.07$.
Hence, it is possible for the junction to exhibit opposite temperature dependences in the
troughs ($\nu\approx 0$) and peaks ($\nu\approx  \frac{1}{2}$) of the conductance as the gate
voltage is varied.  This result is to be contrasted with the behavior at $ \epsilon=0.1$, where
the conductance at $\nu=0.5$ is very weakly temperature dependent.  Indeed, at $ \epsilon=0$
and $\nu= \frac{1}{2}$, the inverse charging energy is known to diverge very weakly\cite{mat95}.
While the experiments of Joyez  {\it et al.\/}\cite{joy97} seem to be perfectly consistent with the
more familiar $ \epsilon=0$ behavior, anomalous temperature
dependence in the troughs has been observed in the conductance of quantum dots\cite{mau99},
although it is hardly clear that the nonequilibrium effects which we consider are uniquely
responsible for this phenomenon.

\section{Conclusions}\label{Scon}
Nonequilibrium shakeup processes have the potential to drastically affect the physics
of tunnel junction behavior, by allowing for a conducting phase in which the Coulomb
blockade is completely suppressed. The phase transition is made manifest within a
large-$N$ approach, although it was first predicted over 25 years ago by Kosterlitz\cite{kos77}
in his renormalization group studies of $ {\rm O} (N)$ classical spin chains with long-ranged
(power law) interactions.

The phase transition has two principal signals.  First, the renormalized charging energy
$E_ {\rm c}^*$ vanishes for $\alpha \ge\alpha_ {\rm c}$.  Second, while the conductance
vanishes for $\alpha < \alpha_ {\rm c}$ and diverges for $\alpha > \alpha_ {\rm c}$, precisely
at the transition $g_ {\rm c}=g(\alpha_ {\rm c})$ is finite and universal (although $ \epsilon$-dependent).
Extracting the critical value $\alpha_ {\rm c}$ from numerical data at finite temperature can be
tricky, we have found.  In particular, the renormalized charging energy shows little signal at
$\alpha_ {\rm c}$ even at dimensionless inverse temperatures on the order of $L=E_ {\rm c}/ k_{\scriptscriptstyle {\rm B}}T
\sim 40$; this behavior is borne out explicitly in our large-$N$ studies.  It is more
reliable to obtain $\alpha_ {\rm c}$ from the crossing of the conductance curves $g(\alpha,L)$.

An extension of the model investigated here to the case of granular systems has been
recently considered\cite{aro02}.

\section{Acknowledgements}\label{Sack}
We thank F. Guinea, C. Herrero, and E. Bascones for fruitful discussions.
We are particularly grateful to F. Guinea for his critical reading of our
manuscript.

\section{Appendix: Absence of Long-Ranged Order for $ \epsilon=0$}\label{APP}

Applying Mermin's classic Bogoliubov inequality arguments \cite{mer67},
{\v S}im{\'a}nek \cite{sim87} has proven the absence of long-ranged order
in the $1/n^2$ XY chain.  Here we (trivially) extend {\v S}im{\'a}nek's
work to the continuum, and show how the $ \epsilon=0$ case is marginal,
 {\it i.e.\/}\ absence of order cannot be so proven for $ \epsilon>0$.

We begin with an action
\begin{eqnarray}
&&S[P(s),\phi(s)]= \int\limits_0^L\!\! ds\,\left[ \frac{1}{2} P^2(s) + \frac{1}{4} ( {\partial}_s\phi)^2
-h\cos\phi(s)\right]\nonumber\\
&&\quad+\ \alpha \int\limits_0^L\!\! ds \int\limits_0^L\!\! ds'\,K(s-s')\,\big(1-\cos[\phi(s)-\phi(s')]\big)\ .
\end{eqnarray}
The Poisson bracket of $A$ and $B$ is defined by
\begin{equation}
 \left\{ A,B \right\}\equiv \int\limits_0^L\!\! ds\,\left({\delta A\over\delta P(s)}\,
{\delta B\over\delta\phi(s)} - {\delta A\over\delta \phi(s)}\,
{\delta B\over\delta P(s)}\right)\ ,
\end{equation}
and the average of a functional $A[P(s),\phi(s)]$ is
\begin{equation}
\langle A\rangle = {\int\!  {\cal D} P\int\!  {\cal D}\phi\, A[P,\phi]\,
e^{-S[P,\phi]}\over
\int\!  {\cal D} P\int\!  {\cal D}\phi\,e^{-S[P,\phi]}}\ .
\end{equation}
Note that when $A$ is a functional of $\phi(s)$ alone, the functional
integral over the momentum field $P(s)$ cancels between numerator and
denominator,  {\it i.e.\/}\ we recover the modified BMS model.

The Bogoliubov inequality guarantees \cite{mer67}
\begin{equation}
\langle A^* A\rangle\ge {\big|\langle \left\{ C,A^* \right\}\rangle\big|^2\over
\langle \left\{ C, \left\{ C^*,S \right\} \right\}\rangle}\ ,
\end{equation}
for an arbitrary functional $C[P,\phi]$.  Following {\v S}im{\'a}nek, we take
\begin{eqnarray}
A[P,\phi]&=& \int\limits_0^L\!\! ds\,\sin\phi(s)\,e^{-i\omega_n s}\nonumber\\
C[P,\phi]&=& \int\limits_0^L\!\! ds\,P(s)\,e^{-i\omega_n s}\ ,
\end{eqnarray}
where $\omega_n=2\pi n/L$ is a Matsubara frequency.  In the $L\to\infty$
limit, the Bogoliubov inequality then establishes the following identity:
\begin{equation}
\int\limits_{-\infty}^\infty\!\!{d\omega\over 2\pi}\,
{m^2\over hm +  \frac{1}{2}\omega^2 + 2\alpha C_ \epsilon |\omega|^{1- \epsilon}}\le 1\ ,
\end{equation}
where $m=\langle\cos\phi(0)\rangle$ is the average magnetization density.
When $ \epsilon=0$, the integral diverges as $-m^2 \ln (hm)$, hence the inequality
demands $m=0$ when $h=0$,  {\it i.e.\/}\ no long-ranged order.  However, the case
$ \epsilon=0$ is marginal, and for any $ \epsilon>0$ the integral is convergent
when $h=0$.  While this does not rigorously prove the existence
of an ordered phase for $ \epsilon>0$, it is at least consistent with our findings.


\begin{thebibliography}{99}
\bibitem{htc} The dimensionless (units of $e^2/h$) high temperature conductance is given by $g ^{\vphantom{\dagger}}_\infty=4\pi |t|^2\rho ^{\vphantom{\dagger}}_ {\rm L}\rho ^{\vphantom{\dagger}}_ {\rm R}$, where $t$ is the hopping
amplitude across the junction, and $\rho ^{\vphantom{\dagger}}_{ {\rm L}, {\rm R}}$ are the densities of states
on the left and right plates, respectively.
\bibitem{cfoot} The expression of (\ref{kubo}) corresponds to the paramagnetic
part of the conductance.  The diamagnetic part, which we do not consider here,
is assumed not to alter the critical behavior.  An extra factor of $2\pi$ appears as well,
since we are computing the dimensionless conductance, whose units are $e^2/h$.
\bibitem{spurious} We believe that the impressive agreement between the extrapolated
$\alpha_ {\rm c}$ and the function $2/\pi^2 \epsilon$, shown in fig. 3 of ref. \onlinecite{bas00}, is
spurious.  The one-loop renormalization group calculation of Kosterlitz \cite{kos77} predicts
$\alpha_ {\rm c}^{\rm RG}=1/2\pi^2 \epsilon+ {\cal O}$.
\end{thebibliography}
\end{document}